%% file: ms.tex
\documentclass{emulateapj}
\usepackage{psfig}

\newcommand{\fig}[1]{Figure~\ref{#1}}
\newcommand{\tab}[1]{Table~\ref{#1}}

\newcommand{\sex}{\texttt{SExtractor}}
\newcommand{\hst}{HST}

\newcommand{\mlo}{17.5}                 
\newcommand{\mhi}{25.5}                 
\newcommand{\ratio}{0.7}                

\newcommand{\nltd}{17}                  
\newcommand{\nfield}{40}                
\newcommand{\npar}{39}                  
\newcommand{\apar}{185.41}              
\newcommand{\ersarea}{46.49}            
\newcommand{\area}{231.90}              
\newcommand{\npoint}{5982}              

\newcommand{\zmin}{240}                 
\newcommand{\zmax}{350}                 
\newcommand{\deltaz}{5}                 

\newcommand{\zscl}{290}                 
\newcommand{\dzrand}{25}                
\newcommand{\dzdust}{5}                 
\newcommand{\dzlumf}{30}                
\newcommand{\dzsys}{31}                 
\newcommand{\dztot}{40}                 

\shorttitle{High-Latitude Ultracool Dwarfs}
\shortauthors{Ryan Jr. et al.}

\begin{document}

\title{{\it Hubble Space Telescope} Observations of Field Ultracool Dwarfs at High Galactic Latitude\footnote{Based on observations made with the NASA/ESA Hubble Space Telescope, obtained from the Data Archive at the Space Telescope Science Institute, which is operated by the Association of Universities for Research in Astronomy, Inc., under NASA contract NAS 5-26555.}}

\author{R. E. Ryan Jr.\altaffilmark{2},
P. A. Thorman\altaffilmark{2},
H. Yan\altaffilmark{3},
X. Fan\altaffilmark{4},
L. Yan\altaffilmark{5},
M. R. Mechtley\altaffilmark{6},
N. P. Hathi\altaffilmark{7},
S. H. Cohen\altaffilmark{6},
R. A. Windhorst\altaffilmark{6},
P. J. McCarthy\altaffilmark{7}, and D. M. Wittman\altaffilmark{2}}

\email{rryan@physics.ucdavis.edu}

\altaffiltext{2}{Physics Department, University of California, Davis, CA 95616}
\altaffiltext{3}{Center for Cosmology and Astroparticle Physics, Ohio State University, Columbus, OH 43210}
\altaffiltext{4}{Steward Observatory, Tucson, AZ 85721} 
\altaffiltext{5}{Spitzer Science Center, California Institute of Technology, MS220-6, Pasadena, CA 91125}
\altaffiltext{6}{School of Earth and Space Exploration, Arizona State University, Tempe AZ 85287}
\altaffiltext{7}{Observatories of the Carnegie Institute of Washington, Pasadena, CA 91101}

\begin{abstract}

We  present  a  sample   of  \nltd~newly  discovered  ultracool  dwarf
candidates later than $\sim\!\mathrm{M}8$, drawn from \area~arcmin$^2$
of {\it Hubble Space Telescope}  Wide Field Camera 3 infrared imaging.
By      comparing     the      observed     number      counts     for
$\mlo\!\leq\!J_{125}\!\leq\!\mhi$~AB mag to an exponential disk model,
we      estimate       a      vertical      scale       height      of
$z_{scl}\!=\!\zscl\pm\dzrand\;(\mathrm{random})\pm\dzsys\;
(\mathrm{systematic})$~pc  for  a  binarity fraction  of  $f_b\!=\!0$.
While our  estimate is roughly  consistent with published  results, we
suggest that  the differences can be attributed  to sample properties,
with the  present sample containing  far more substellar  objects than
previous  work.    We  predict  the  object  counts   should  peak  at
$J_{125}\!\sim\!24$~AB mag  due to the  exponentially-declining number
density at the edge of the  disc. We conclude by arguing that trend in
scale height with  spectral type may breakdown for  brown dwarfs since
they do not settle onto the main sequence.

\end{abstract}

\keywords{Keywords: Galaxy: structure --- Galaxy: stellar content --- stars: low-mass, brown dwarfs}

\section{Introduction} \label{intro}

Star  counts have long  been used  to determine  the structure  of our
Galaxy.   Early attempts  were  plagued by  patchy  extinction and  by
mathematical   instabilities   in  the   inversion   of  star   counts
\citep[see][]{bok37},  but \citet{bs80,bs84}  revived the  endeavor by
avoiding  regions  with  significant  extinction,  and  by  fitting  a
physically  motivated model with  only a  few parameters  (see Bahcall
1986 for a review). Recently,  there has been renewed interest in star
counts as  ever-cooler stellar populations are discovered  and need to
be                                                              modeled
\citep{ryan05,nor05,caba08,juric08,nor09,deacon09,boch10,del10}.
Because brown dwarfs cool and change spectral type on relatively short
timescales, their vertical scale heights may reflect not just Galactic
structure, but also their cooling times.

The  Galactic  distribution  of  the ultracool  dwarf  population  has
garnered  much interest  from a  community studying  far  more distant
objects.  Since the extremely  red optical and near-infrared colors of
the  ultracool dwarfs  are similar  to those  of  Lyman-break galaxies
\citep[LBGs;][]{steid96}           at           redshifts           of
$5\!\lesssim\!z\!\lesssim\!7$,    there   are   concerns    that   the
high-redshift  galaxy samples  may be  contaminated by  these Galactic
objects  \citep[e.g.][]{caba08}.   In  the  absence  of  spectroscopic
identification, LBG  studies often resort  to statistically correcting
their      number     counts     for      foreground     contamination
\citep[e.g.][]{bouwens06}.   Naturally,   this  correction  relies  on
accurately  characterizing  the  Galactic  distribution  of  ultracool
dwarfs.    To  this  end,   \citet{ryan05}  identify   28~dwarfs  with
$(i'-z')_{\rm  AB}\!\geq\!1.3$~mag \citep[which  are types  later than
  $\sim\!\mathrm{M}6$;][]{boch10}  in   15~parallel  fields  from  the
Advanced  Camera  for  Surveys  (ACS)  aboard the  {\it  Hubble  Space
  Telescope}   (\hst).   By  assuming   an  exponential   disk  model,
\citet{ryan05}     derive    a     vertical     scale    height     of
$z_{scl}\!=\!350\pm50$~pc,  and conclude that  the deepest  surveys of
$z\!\simeq\!6$    LBGs   were   $\gtrsim\!97$\%    pure.    Similarly,
\citet{nor05}     find     a      vertical     scale     height     of
$z_{scl}\!=\!400\pm100$~pc  from  three  spectroscopically  identified
late-M   and  early-L   dwarfs  in   the  Hubble   Ultra   Deep  Field
\citep{beck06}.

The overwhelming majority of ultracool  dwarfs to date have been found
in          shallow,          very         wide-field          surveys
\citep[e.g.][]{delf99,kirk99,knapp04},  and more recently  with deeper
datasets  \citep[e.g.][]{del08,deacon09}.   Since  these  objects  are
intrinsically  very faint \citep[$M_{i'}\!\gtrsim\!16$~mag;][]{haw02},
nearly all known ultracool  dwarfs reside within $\sim\!100$~pc of the
Sun \citep[e.g.][]{reid08}, which makes determining the Galactic-scale
distribution  difficult  or  impossible.   While  this  issue  can  be
mitigated to a large extent  by probing further into the disk, limited
observing  time and  detection efficiency  have restricted  studies to
narrow      fields-of-view      and/or      single      lines-of-sight
\citep[e.g.][]{ryan05,nor05,nor09}.     Naturally   this    leads   to
simplified  models,  large  uncertainties  on  model  parameters,  and
significant variations  between authors.  In  this paper, we  begin to
overcome  these  limitations by  drawing  our  sample  from very  deep
parallel and pointed  fields with \hst, which have  the sensitivity to
find an L0-dwarf to  $\sim\!3.5$~kpc and a T0-dwarf to $\sim\!700$~pc.
These  represent  a  significant  increase  in  survey  distances  for
ultracool dwarfs.

This work  is organized  as follows: in  \S~\ref{obs} we  describe the
observations and  source catalogs,  in \S~\ref{sample} we  discuss our
ultracool dwarf sample selection,  in \S~\ref{analysis} we present our
analysis and  scale height  measurement, in \S~\ref{sysunc}  we assess
our systematic uncertainties, and  in \S~\ref{disc} we conclude with a
brief  review   and  thoughts  toward   future  improvements.   Unless
explicitly stated  otherwise, all magnitudes  and colors are  given in
the AB system \citep{abmag}.

\section{Observations} \label{obs}

Here we discuss the parallel and pointed fields with \hst\ which 
consititute our dataset.  We list their salient properties in \tab{fields}.

\subsection{The \hst\ Parallel Imaging} \label{purepars}

The bulk  of the data analyzed  here come from  the high-level science
products from the Hubble  Infrared Pure Parallel Imaging Extragalactic
Survey
\citep[HIPPIES\footnote{http://archive.stsci.edu/prepds/hippies/};][]{yan11}.
At present,  this survey combines  two pure parallel  imaging programs
with \hst\  Wide Field Camera~3  (WFC3; PropIDs: 11700 and  11702) and
coordinated parallels\footnote{See  the \hst\ User  Information Report
  UIR-2008-001 for a discussion  of parallel imaging with \hst.}  from
the  Cosmic Origins  Spectrograph (COS)  guaranteed  time observations
(GTO).  Every field  has infrared imaging in F098M,  F125W, and F160W,
and  optical imaging  in  F606W and/or  F600LP\footnote{The fields  at
  $02^{\rm  h}20^{\rm  m}$,   $07^{\rm  h}50^{\rm  m}$,  and  $12^{\rm
    h}09^{\rm m}$ have both  optical bands.}.  By design, these fields
are at relatively  high Galactic latitude ($|b|\!>\!20^{\circ}$), have
a  total exposure time  of $\geq\!4$~\hst\  orbits, and  sample random
pointings through the Galaxy.

The data reduction and mosaicking of the HIPPIES data are discussed in
detail by  \citet{yan11}, but we  will mention the key  steps relevant
for this  work. Standard procedures were followed  except for enhanced
removal of  image defects.  Since  the \hst\ parallel data  are rarely
dithered,  the affected  pixels were  corrected by  interpolating over
neighboring pixels with the \texttt{FIXPIX} routine in IRAF.  The main
side-effect of this  procedure is to decrease the  usable area of each
parallel pointing.  In total,  we analyzed \npar~parallel fields which
cover \apar~arcmin$^2$.   Finally, we note  that the COS  GTO parallel
fields also have additional imaging in F300X, F475W, and F475X, though
we place no constraints on the colors in these bands.

\subsection{The WFC3 Early Release Science Program} \label{ers}

In addition  to the parallel data,  we include the  WFC3 Early Release
Science program (ERS; PropID:~11359)  data taken in the southern field
of     the     Great     Observatories     Origins     Deep     Survey
\citep[GOODS-S;][]{giav04}.  The WFC3 imaging in the F098M, F125W, and
F160W-bands augments the existing optical data from the ACS and covers
$\ersarea$~arcmin$^2$.  The ERS data have at least double the exposure
time of the WFC3 parallel data  in all bands. Details of the ERS data,
such  as experimental  design, reduction,  and imaging  properties are
discussed  by \citet{win11}.   Like  the COS  GTO  parallels, the  ERS
subset  of the  GOODS-S field  has  been observed  in many  additional
optical and ultraviolet bands,  however we will not impose constraints
on those colors to ensure a uniformly selected sample.

\input{tab1.tex}

\subsection{Photometry} \label{sexcat}

We    measure    all    magnitudes    as    \texttt{MAG\_AUTO}    with
\sex\  \citep{bert}  in dual-image  mode  using  the  F125W image  for
detection.   We  require  a  minimum  area of  5~connected  pixels,  a
threshold (per pixel) of $\geq\!1.5~\sigma$ over the local background,
and  use a  $5\times5$~pix Gaussian  filter  with full  width at  half
maximum of $2$~pix  for source detection.  All images  are drizzled to
$0\farcs090$~pix$^{-1}$.   We   use  64~deblending  sub-thresholds,  a
minimum contrast parameter of  $10^{-4}$, and a cleaning efficiency of
10.   We adopt the  AB zeropoints  from \citet{kali30,kali31}  for the
WFC3          data          of         $\mathrm{F606W}\!=\!26.08$~mag,
$\mathrm{F600LP}\!=\!25.85$~mag,        $\mathrm{F098M}\!=\!25.68$~mag,
$\mathrm{F125W}\!=\!26.27$~mag,   and  $\mathrm{F160W}\!=\!25.96$~mag,
and $\mathrm{F606W}\!=\!26.486$~mag  for the  ACS data in  the GOODS-S
field.   Based   on  these  object  catalogs,  we   compute  the  50\%
completeness depth in the F125W image as the magnitude where the ratio
of the  observed counts to  a power-law fit  falls to 0.5,  and report
these depths in \tab{fields}.

\begin{figure}
\epsscale{1.2}
\plotone{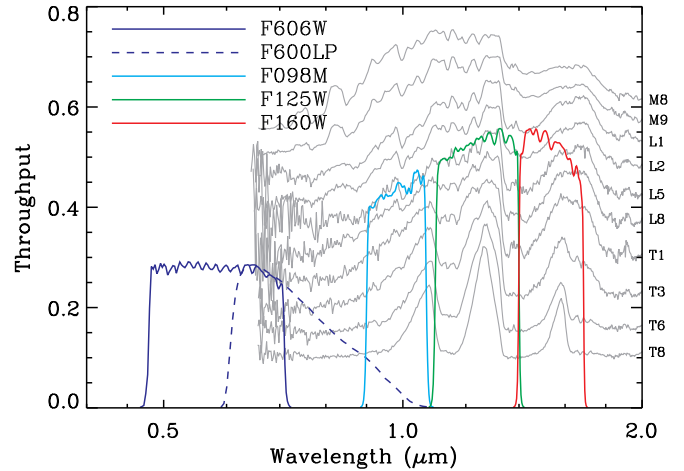}
\caption{\hst/WFC3  filter throughput curves.   The five  filters used
  here are shown as  F606W ($V_{606}$; solid blue), F600LP ($I_{600}$;
  dotted blue), F098M ($Y_{098}$; solid cyan), F125W ($J_{125}$; solid
  green), and F160W  ($H_{160}$; solid red).  In light  gray lines, we
  show  select spectra  from  the A.~Burgasser  SpeX compilation  (see
  \tab{spexcat}) with  infrared types indicated on the  left.  The COS
  coordinated parallels and the ERS field have additional blue filters
  from \hst. However, we do  not explicitly place constraints on these
  colors, in order to ensure a uniform sample.\label{filters}}
\end{figure}

The parallel  and ERS  data give us  a total  of $\nfield$~independent
sight   lines   through   the    disk   which   cover   a   total   of
$\area$~arcmin$^2$.  As mentioned above, each field has the same three
near-infrared   bands;  however,  the   optical  imaging   differs  in
wavelength and instrument.  In  \fig{filters}, we show our optical and
infrared bandpasses:  F606W (solid blue), F600LP  (dashed blue), F098M
(cyan),  F125W  (green),  and  F160W  (red),  which  we  refer  to  as
$V_{606}$,    $I_{600}$,   $Y_{098}$,   $J_{125}$,    and   $H_{160}$,
respectively.

\input{tab2.tex}

\section{Ultracool Dwarf Candidates} \label{sample}
\subsection{Sample Selection}\label{sampleselect}

\begin{figure}
\epsscale{1.2}
\plotone{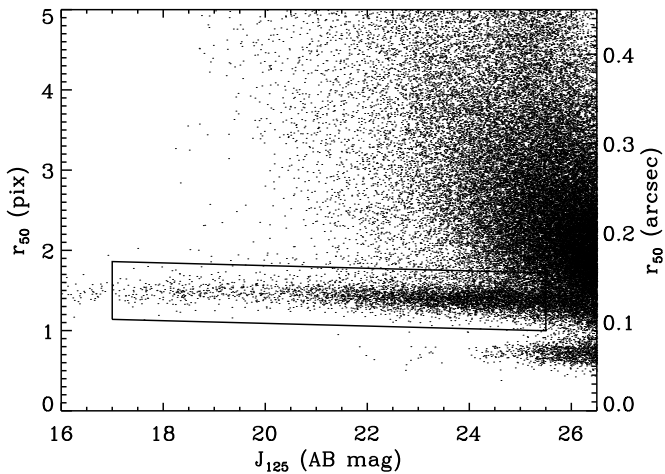}
\caption{Morphological  selection  criteria.   We use  the  half-light
  radius,    axis    ratio,   and    brightness    as   measured    by
  \texttt{SExtractor}  to  identify  point  sources.  The  small  dots
  represent all  objects from the \area~arcmin$^2$  analyzed here, and
  the solid lines show the  stellar locus selection region.  We find a
  very weak  relationship between half-light radius  and brightness.
  Our constraints  on half-light radius and axis  ratio are consistent
  with   known   stars   selected   from   the   Sloan   Digital   Sky
  Survey.\label{locus}}
\end{figure}

To ensure that  our objects are point-like, we  require the axis ratio
to be  $(b/a)\!\geq\!\ratio$ and the half-light radius  as measured by
\texttt{FLUX\_RADIUS}                                                in
\texttt{SExtractor}\footnote{\texttt{SExtractor}   will   measure  the
  radius at which  a some fraction of the total  flux is reached based
  on the  setting $\texttt{PHOT\_FLUXFRAC}$,  which we adopt  as 0.5.}
to    be   $1.2\!\leq\!r_{50}\!\leq\!1.8$~pix.     While   unsaturated
point-sources  should  have half-light  radii  roughly independent  of
brightness, we find  a weak trend in the  stellar locus, therefore our
exact half-light radii limits vary slightly with magnitude.  Using the
morphological   criteria  presented   in   \fig{locus},  we   identify
\npoint~point sources in the \area~arcmin$^2$ surveyed.

\begin{figure}
\epsscale{1.2}
\plotone{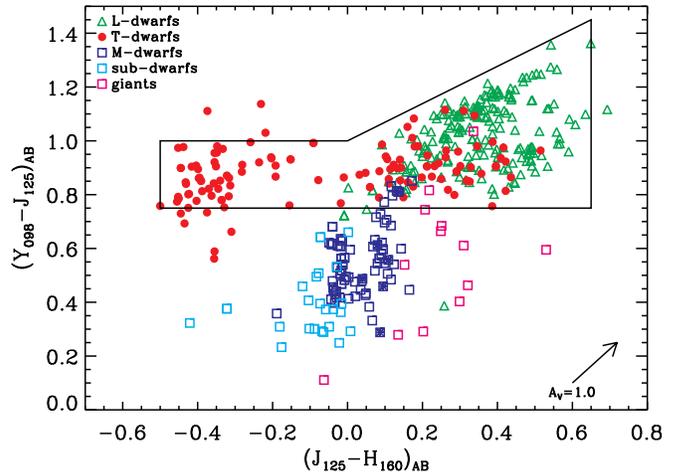}
\caption{Color--color diagram for  ultracool dwarf selection.  We show
  library  of empirical  L- and  T-dwarfs from  the  SpeX spectrograph
  compiled by A.~Burgasser (see  \tab{spexcat}) as green triangles and
  red  circles,  respectively.   We  show the  M-dwarfs,  giants,  and
  subdwarfs as  blue, magenta, and cyan  squares, respectively.  Based
  on   these   colors,   we    define   unique   spectral   types   as
  equations~(\ref{mltdwarf}),            (\ref{ldwarf}),           and
  (\ref{tdwarf}). \label{modcol}}
\end{figure}

We select  our ultracool  dwarf candidates from  the catalog  of point
sources  based   on  their  optical  and   near-infrared  colors.   We
synthesize    empirical    $(V_{606}-Y_{098})$,   $(I_{600}-Y_{098})$,
$(Y_{098}-J_{125})$, and $(J_{125}-H_{160})$  colors from a library of
spectra from the 3-meter  NASA Infrared Telescope Facility compiled by
A.~Burgasser\footnote{Distributed                                    at
  http://web.mit.edu/ajb/www/browndwarfs/spexprism/.}     (listed   in
\tab{spexcat}).   In \fig{modcol}, we  show the  infrared color--color
diagram for known L-dwarfs  (green triangles), T-dwarfs (red circles),
and M-dwarfs, giants, and  subdwarfs (blue, magenta, and cyan squares,
respectively).   Unfortunately, our  broadband data  cannot accurately
constrain the  spectral type since the  $J_{125}$- and $H_{160}$-bands
equally sample the strong H$_2$O absorption at $\lambda\!=\!1.34~\mu$m
(see \fig{filters}).  Therefore, the  only unique spectral types we can
derive from these near-infrared data are given by

\begin{eqnarray} \label{mltdwarf}
  \mathrm{MLT}\left\{\begin{array}{ll}
  0.0\leq(J_{125}-H_{160})\leq0.65 \:\hbox{mag; and}\\
  0.75\leq(Y_{098}-J_{125})\leq1.0\:\hbox{mag}
  \end{array}\right.
\end{eqnarray}

\begin{eqnarray} \label{ldwarf}
  \mathrm{L}\left\{\begin{array}{ll}
   0.0\leq(J_{125}-H_{160})\leq0.65 \:\hbox{mag; and}\\
   1.0\leq(Y_{098}-J_{160})\leq0.7\times(J_{125}-H_{160})+1.0\:\hbox{mag}
  \end{array}\right.
\end{eqnarray}

\begin{eqnarray} \label{tdwarf}
  \mathrm{T}\left\{\begin{array}{ll}
  -0.5\leq(J_{125}-H_{160})\leq0.0 \:\hbox{mag; and}\\
  0.75\leq(Y_{125}-J_{160})\leq1.0\:\hbox{mag}
  \end{array}\right.
\end{eqnarray}
For  all types,  we require  $(V_{606}-Y_{098})\!\geq\!2.0$~mag and/or
$(I_{600}-Y_{098})\!\geq\!0.5$~mag.   We present  our  ultracool dwarf
candidates in \tab{cands}.

\begin{figure}
\epsscale{1.2}
\plotone{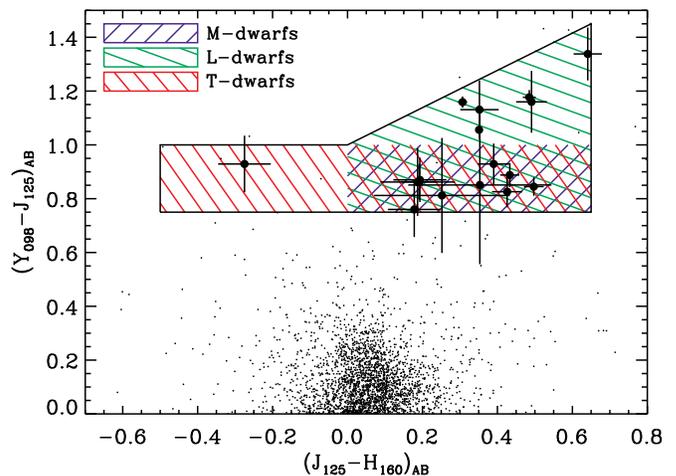}
\caption{Color--color diagram for  ultracool dwarf candidates. We show
  the   selection   region   defined  by   equations~(\ref{mltdwarf}),
  (\ref{ldwarf}), and  (\ref{tdwarf}) as  thick lines and  all objects
  passing our  brightness and morphological criteria  as small points.
  In blue, green, and red hatches,  we show the expected colors of M-,
  L-, and T-dwarfs to highlight the ambiguity in spectral typing these
  stars  with   these  bandpasses.   The   points  with  uncertainties
  represent     our     \nltd~ultracool     dwarf    candidates     in
  \tab{cands}.\label{obscol}}
\end{figure}

\subsection{Contaminants} \label{contaminants}

In addition to the ultracool  dwarfs, there are three additional types
of  known astrophysical  objects  which may  satisfy  our optical  and
near-infrared  color criteria: early-type  galaxies (ETGs),  LBGs, and
high redshift quasars.  In the absence of spectroscopic confirmation, we 
can only make statistical arguments on these potential objects:

Based on the \citet{cww80} spectral templates, we estimate that an ETG
at  $1.2\lesssim\!z\!\lesssim\!1.5$  will  have optical  and  infrared
colors  similar  to  our   ultracool  dwarfs.   By  extrapolating  the
luminosity function of ETGs at $z\!=\!1$ \citep{faber07}, we find that
our fields  could have  $\sim\!80$~ETGs at these  redshifts.  However,
the number  of ETG  contaminants in our  sample should be  much lower,
since we  require our stellar  candidates be unresolved, which  is not
represented in  this simple  brightness calculation.  To  estimate the
fraction of these  ETGs that are also unresolved,  we perform a simple
Monte  Carlo simulation.   We draw  $10^5$ random  absolute magnitudes
from the $z\!=\!1$ ETG  luminosity function over our sample brightness
range, which  we convert to stellar masses  assuming the mass-to-light
of  $\Upsilon_B\!=\!1~M_{\odot}~L_{\odot}^{-1}$.  For a  given stellar
mass, we draw  a random size according to  the mass--size relation for
local ETGs  from the SDSS  \citep{shen03}, and determine  the measured
effective   radius   by  quadratically   adding   the   size  of   the
$J_{125}$-band  PSF  $r_{\rm  meas}\!=\!\sqrt{r_{\rm  SDSS}^2+r_J^2}$.
Finally,  we take  the fraction  of deviates  which satisify  our size
criterion (see  \fig{locus}) as the fraction of  detectable ETGs which
would be unresolved in these  \hst\ images.  We estimate the potential
ETG fraction to be  $\lesssim\!0.1$\% for the \citet{shen03} relation,
and this  fraction only rises to $\lesssim\!1$\%  for the $z\!\sim\!2$
mass--size  relations \citep[e.g.][]{ryan10}.  Therefore,  we conclude
that our sample is largely free of contaminating ETGs.

Like the ETGs, LBGs and quasars  can only corrupt our sample in a very
specific  redshift range  of  $6.8\!\lesssim\!z\!\lesssim\!7.2$. While
both populations are  likely to be unresolved, LBGs  are typically far
too faint, and quasars are far  too rare, to have been included in our
sample.            In          our           brightness          range
($\mlo\!\leq\!J_{125}\!\leq\!\mhi$~mag),    we    expect    to    find
$\sim\!0.02$~LBGs     and    $\sim\!0.01$~quasars,     assuming    the
\citet{bouwens10}    and    \citet{will10}    luminosity    functions,
respectively.  Therefore, we conclude our sample is likely free of any
LBGs and/or quasars.

\input{tab3.tex}

\section{Analysis} \label{analysis}

\begin{figure}
\epsscale{1.2}
\plotone{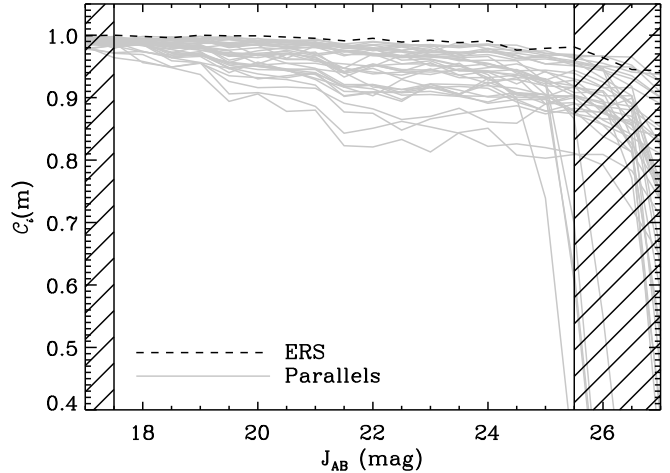}
\caption{The  completeness corrections for  the parallel  (solid gray)
  and ERS field (dotted black).   We estimate these corrections as the
  recovery  rate  of  1000   randomly-placed  point  sources  in  each
  magnitude  interval.  The  completeness  corrections  are  generally
  $\gtrsim\!90\%$  for our magnitude  range and  rarely tend  to 100\%
  (even at the bright-end), due to source blending.\label{compfig}}
\end{figure}

We  determine  the  vertical  scale  height  of  ultracool  dwarfs  by
comparing our observed star counts  to those predicted from a Galactic
structure  model.  We  model  the spatial  distribution  of dwarfs  as
\begin{equation}
n(r,z)\!\propto\!e^{-(r-r_{\odot})/r_{scl}}e^{-|z|/z_{scl}},
\end{equation}
where $r_{\odot}\!=\!8$~kpc is the Solar position, $r_{scl}\!=\!2$~kpc
is  the  radial scale  length  \citep{juric08},  and  the constant  of
proportionality  is  set by  the  local ($R\!\leq\!20$~pc)  luminosity
function, $\Phi(M)$.   We take  the empirical luminosity  functions of
\citet{cruz07}  for  the  M8--L9-dwarfs,  and of  \citet{reyle10}  for
T-dwarfs.  The model number counts for the $i$th field are
\begin{equation}\label{num}
\hat{N}_m(\ell_i,b_i)\,dm=\Delta\Omega_i\,\mathcal{C}_i(m)\,dm\int_0^{\infty}\!R^2n(r_i)\Phi(M)\,dR,
\end{equation}
where ($\ell_i,b_i$) are the Galactic coordinates, $\Delta\Omega_i$ is
the                solid                angle               subtended,
$M\!=\!m-5\log\left(R\right)-5-A\left(\ell_i,b_i,R\right)$    is   the
absolute  magnitude,  $R$ is  the  heliocentric  distance in  parsecs,
$A(\ell_i,b_i,R)$  is  the extinction  (discussed  in  more detail  in
\S~\ref{dust}),                    $x_i\!=\!\sqrt{r_{\odot}^2+R^2\cos^2
  b_i-2Rr_{\odot}\cos\ell_i\cos  b_i}$  is   the  distance  along  the
Galactic  midplane,  $z_i\!=\!R\sin b_i$  is  the  distance above  the
Galactic midplane,  $r_i\!=\!\sqrt{x_i^2+z_i^2}$ is the Galactocentric
distance \citep{bahc86},  and $\mathcal{C}_i(m)$ is  the completeness.
The total  model number counts  are given by  the sum over all  of the
fields
\begin{equation}
\hat{N}_m=\sum_{i=1}^{N_{\rm fields}}\hat{N}_m(\ell_i,b_i),
\end{equation}
which is  parameterized by the  vertical scale height in  the Galactic
model.   We estimate the  magnitude-dependent completeness  by placing
point-sources  of known  brightness  at random  locations within  each
field,  cataloging the  images  as discussed  in \S~\ref{sexcat},  and
taking the completeness as the  fraction of recovered objects. In this
way, we  encapsulate the effects of  our choice of  \sex\ settings and
source   blending.   In  \fig{compfig},   we  show   the  completeness
corrections for  each parallel  field (light gray  lines) and  the ERS
field (dashed black line).  The hatched regions indicate the magnitude
ranges  that   we  omit  in   our  analysis.   Given   our  relatively
conservative  limits  of $\mlo\!\leq\!J_{125}\!\leq\!\mhi$~mag,  these
completeness values are rarely  $\lesssim\!90$\%, and generally do not
fall to their half-maximum values until $J_{125}\!\simeq\!26$~mag.

\begin{figure}
\epsscale{1.2}
\plotone{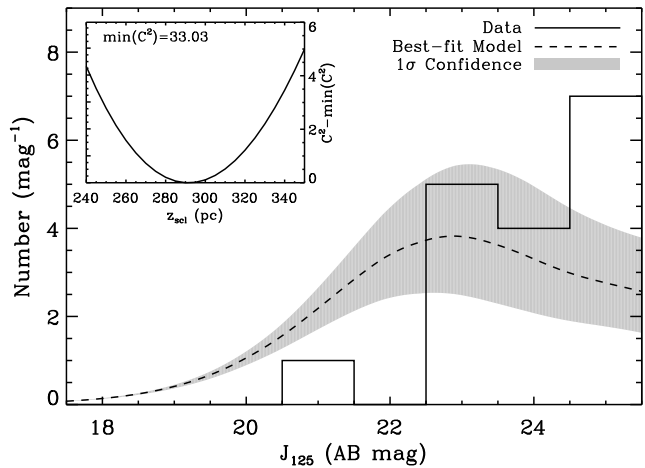}
\caption{Ultracool dwarf number  counts.  The histogram represents the
  number counts from \area~arcmin$^2$  from \npar\ parallel fields and
  the ERS data  in the GOODS-S field.  The solid  line and grey region
  shows best-fitting  model counts for  $z_{scl}\!=\!\zscl$~pc and the
  corresponding  $1\sigma$ uncertainty,  respectively.   In the  upper
  left, we show  the $\Delta C^2(z_{scl})$ used to  derive these scale
  height   values.   The   peak  in   the  model   number   counts  at
  $J_{125}\!\sim\!24$~mag is not  due to generic object incompleteness
  at  the  faint-end, but  rather  from  having integrated  completely
  through the disk.\label{counts}}
\end{figure}

We      compute       the      model      number       counts      for
$\zmin\!\leq\!z_{scl}\!\leq\!\zmax$~pc           with          $\Delta
z_{scl}\!=\!\deltaz$~pc and  maximize the likelihood  of obtaining the
measured counts.  Since the observed  counts are in the limit of small
integers, the probability distribution  should be modeled as a Poisson
distribution  \citep{cash79}.   To  avoid  confusion  with  the  usual
Gaussian  probability distributions,  we  denote this  goodness-of-fit
statistic as  $C^2$, and maximize  the likelihood (${\cal L}$)  in the
usual way:
\begin{eqnarray}
C^2(z_{scl})&=&-2\ln\left(\prod_{m}{\cal L}\left(N_m\left|\hat{N}_m\right.\right)\right)\\
&=&-2\sum_{m}N_m\ln\left(\hat{N}_m\right)-\hat{N}_m-\ln\left(N_m!\right)
\end{eqnarray}
where $N_m$  and $\hat{N}_m$ are  the observed and  model differential
number   counts,    respectively.    Finally,   we    define   $\Delta
C^2\!\equiv\!C^2-\min(C^2)$,   which    will   follow   the   standard
$\chi^2$-distribution  \citep{cash79} with one  degree of  freedom (in
this case  the scale height).   In \fig{counts}, we show  the observed
(solid histogram) and model counts (dashed line) for the optimal model
of  $z_{scl}\!=\!\zscl$ with the  total $1\sigma$  uncertainty (shaded
region    ---   we   discuss    our   systematic    uncertainties   in
\S~\ref{sysunc}),  as  well as  the  $\Delta  C^2(z_{scl})$ curve  for
$A_J\!=\!0$~mag in  the inset (discussed in  detail in \S~\ref{dust}).
We compute  the random uncertainty  on the scale height  where $\Delta
C^2\left(z_{scl}\right)\!=\!1$ to be $\pm\dzrand$~pc.

\section{Systematic Biases and Uncertainties} \label{sysunc}

Here  we discuss  sources  of potential  systematic uncertainties  that 
could affect our  estimate of the vertical scale height.

\subsection{Interstellar Extinction} \label{dust}

As  mentioned above,  it is  necessary to  incorporate the  field- and
distance-dependent extinction  to properly interpret  the star counts.
While the  dust maps of \citet{sfd}  provide the best  estimate of the
total  line-of-sight Galactic  extinction  for extragalactic  objects,
they  cannot be  directly applied  to  our objects  which reside  {\it
  within}  the  Galaxy. Instead  of  parameterizing $A(\ell,b,R)$,  we
perform  the above minimization  for both  $A(\ell,b,R)\!=\!0$~mag and
the  \citet{sfd} value  as given  in \tab{fields}.   As  this approach
brackets  the  two  extinction   extremes  (the  minimum  and  maximum
extinction models, respectively), we  expect it to indicate the degree
to  which   insufficient  knowledge  of   $A(\ell,b,R)$  is  adversely
affecting our  results.  Therefore we take the  average and difference
between the two extremal dust  hypotheses as the expected value of the
scale  height, and its  systematic uncertainty  due to  the extinction
model,   respectively.   We    approximate   this   as   a   symmetric
uncertainty. However  we do not  expect these objects to  be uniformly
distributed between the two extinction limits.  We find that the range
of  scale heights  for the  minimum and  maximum extinction  models is
$\pm\dzdust$~pc.

\subsection{Ultracool Dwarf Luminosity Function}\label{lf}

The  measured  ultracool  dwarf  luminosity function  has  potentially
sizeable  uncertainties ($\delta \Phi/\Phi\!\sim\!30$\%),  largely due
to  the  Poisson  counting  uncertainty  of these  rare  objects.   To
estimate the uncertainty introduced by the inaccurate knowledge of the
luminosity function, we draw a  normal random number for each absolute
magnitude  bin  with mean  and  variance  from  the published  results
\citep[e.g.][]{cruz07,reyle10}.   We   recompute  the  vertical  scale
height for 1000~realizations, and take the dispersion of optimal scale
heights  to   be  the   systematic  uncertainty.   We   estimate  this
uncertainty for our fields to be $\pm\dzlumf$~pc.

\subsection{Malmquist Bias}\label{malm}

For  any   flux-limited  survey,  intrinsically   bright  objects  are
preferentially  detected,  which biases  the  mean absolute  magnitude
($\overline{M}$)  as a  function of  apparent  magnitude \citep{malm}.
Under basic assumptions, the  correction to the absolute magnitudes is
given by
\begin{eqnarray}
\Delta M&=&\overline{M}-M_0,\\
&=&-\frac{\sigma^2}{\log e}\frac{dN_m}{dm},
\end{eqnarray}
where  $M_0$ is  the  intrinsic absolute  magnitude,  $\sigma$ is  the
dispersion on the brightnesses from the width of the main sequence and
photometric    uncertainties    \citep[e.g.][]{boch10},   and    $\log
e\!=\!0.4343$.    Since   our  observed   counts   ($N_m$)  are   very
discontinuous  due  to  small  number  statistics, we  opt  to  impose
Malmquist  bias   on  the   model  counts  ($\hat{N}_m$).    We  adopt
$\sigma\!=\!0.2$~mag  which is a  somewhat more  conservative estimate
than the typical photometric  uncertainty of our faintest sources (see
\tab{cands}),  which  gives  shifts   on  the  absolute  magnitude  of
$-0.1\!\lesssim\!\Delta M\!\lesssim\!0.2$~mag\footnote{Since our model
  counts  peak  at  $J_{125}\!\sim\!24$~mag,  the  absolute  magnitude
  shifts are  not always positive.}. Since this  shift is considerably
smaller than the width of our apparent magnitude bins, the bias on the
vertical scale height is negligible.

\subsection{Equal-Mass Binaries}\label{binaries}

We  expect  a fraction  of  our ultracool  dwarfs  will  be in  binary
systems, which if unaccounted for,  will tend to increase the vertical
scale  height  measurements  \citep[e.g.][]{boch10}.   To  assess  the
properties of potential  binary systems in our sample,  we construct a
grid of simulated  images with two point sources placed  at a range of
separations  ($1\!\leq\!s\!\leq\!5$~pix and  $\Delta s\!=\!0.25$~pix),
total   magnitudes   ($20\!\leq\!J_{125}\!\leq\!25$~mag  and   $\Delta
J_{125}\!=\!0.25$~mag),  and  Gaussian   noise  field  with  mean  and
variance consistent with the parallel fields.  For each brightness and
separation, we  generate 1000~realizations  and catalog the  images as
described  in  \S~\ref{sexcat}.   We  find  that  for  separations  of
$s\!\lesssim\!3$~pix \sex\ does not detect two distinct point sources,
but  does recover  the total  flux to  $\sim\!2$\%.   Furthermore, the
combined  source  only fails  to  pass  our  axis ratio  criterion  of
$(b/a)\!\leq\!\ratio$          for         $J_{125}\!\gtrsim\!23$~mag.
\citet{burgasser2007} find that most very low-mass stars have physical
separations  of  $\Delta\!\lesssim\!20$~AU,  which  implies  that  the
unresolved binaries that  may escape our cataloging are  at a distance
of  $50\!\lesssim\!R\!\lesssim\!80$~pc.   Such  systems will  have  an
absolute magnitude of  $M_J\!\gtrsim\!18$~AB mag, which corresponds to
a spectral  type that  is far too  cool to  have been included  in our
sample \citep{haw02}.  Therefore we  did not systematically reject any
marginally-resolved binaries based on our axis ratio criterion.

Unresolved equal-mass binaries  will be $2.5\log(2)$~mag brighter than
a single star of the same  spectral type and distance, which will skew
the observed  counts to brighter values  and give the  impression of a
thinner  disk  \citep{boch10}.   To  estimate the  magnitude  of  this
effect,  we randomly  select a  fraction  of our  objects (denoted  as
$f_b$)  to   be  equal-mass  binaries.    We  dim  these   objects  by
$2.5\log(2)$~mag, duplicate their entries  in the number counts if the
dimmed  brightness  is  $J_{125}\!\leq\!\mhi$~mag, and  recompute  the
vertical scale height according  to \S~\ref{analysis}.  We repeat this
procedure  1000~times for $f_b\!=\!0.1$,  0.2, 0.3,  and 0.4  and find
that   the   biases   on   the  scale   heights   are   $(z_{scl}^{\rm
  binary}-z_{scl})\!=\!5\pm4$~pc,   $6\pm10$~pc,   $14\pm12$~pc,   and
$17\pm12$~pc, respectively.  The uncertainties in these biases reflect 
the distribution of estimated scale heights.

\section{Discussion} \label{disc}

With  the deep  ($J_{125}\!\lesssim\!26$~mag)  \hst/WFC3 parallel  and
pointed fields, we can  identify an L0-dwarf out to $R\!\sim\!3.5$~kpc
and  a T0-dwarf  to $\sim\!700$~pc.   Since these  fields are  at high
Galactic   latitudes,  they  provide   constraints  on   the  vertical
distribution  of these  intrinsically  very faint  objects.  From  our
compilation of  \area~arcmin$^2$ of \hst\ imaging,  we have identified
\nltd~ultracool dwarf  candidates, whose number  counts are consistent
with  an exponential  vertical  distribution with  a  scale height  of
$z_{scl}\!=\!\zscl\pm\dzrand\;(\mathrm{random})\pm\dzsys\;
(\mathrm{systematic})$~pc.    Our  estimate  improves   upon  previous
results  by combining the  depths of  \citet{nor05} with  the multiple
sight-lines  and  area of  \citet{ryan05}.   Additionally, our  sample
likely contains  fewer M-dwarfs, owing  to the deep  infrared imaging.
For  example, the  \citet{ryan05}  work identified  dwarfs later  than
$\sim\!\mathrm{M}6$  from \hst/ACS  parallels with  a single  color of
$(i'-z')\!\geq\!1.3$~mag \citep[see][for representative SDSS colors of
  M-dwarfs]{boch07}.   Since the  early M-dwarfs  are  of considerably
higher luminosity and far more  common than the L-dwarfs, we speculate
that  the \citet{ryan05}  sample  contains many  M-dwarfs.  Using  our
derived Galactic structure model, we estimate that our sample contains
$6.0\pm2.2$,  $7.4\pm2.2$, and  $2.3\pm0.3$ M8--M9,  L,  and T-dwarfs,
respectively.    Had  we   adopted  the   weaker  color   criteria  of
\citet{ryan05}, we would  expect $27\pm10$~M6--M9 dwarfs.  Given these
likely differences in sample properties,  it is not surprising to find
possible  differences  in  the  vertical  scale  height  measurements.
Finally,    we   note    that   the    model   counts    peak   around
$J_{125}\!\simeq\!24$~mag,  much brighter  than the  50\% completeness
limit of  $J_{125}\!\simeq\!26$~mag (see \tab{fields}).   This peak is
{\it not} due  to generic survey incompleteness at  the faint-end, but
rather  due to  the number  density declining  faster than  the volume
surveyed.   At present,  our observed  number  counts do  not show  or
strongly    demand    such    a    peak,   and    more    dwarfs    at
$J_{125}\!\sim\!24$~mag are needed to identify this critical peak.

We  have  used  the  most  recent estimates  of  the  ultracool  dwarf
luminosity function  \citep[e.g.][]{cruz07,reyle10}, which are derived
primarily from nearby  samples ($R\!\lesssim\!100$~pc).  Since objects
below   the  hydrogen-burning   limit  are   passively-cooling,  their
bolometric  luminosity  strongly  depends  on their  age  and  initial
temperature.  Therefore, the cooling will introduce a non-trivial time
dependence  on the luminosity  function of  a population  of ultracool
dwarfs \citep{burg04lf}.  For example, if the majority of these dwarfs
are  formed at  the  disk midplane  and  are scattered  to these  high
Galactic latitudes  by interactions with massive objects  in the disk,
then the luminosity function of  these dwarfs is likely different than
the  local estimates, particularly  if the  cooling times  are shorter
than  the  scattering  times.   Specifically,  many  of  the  earliest
L-dwarfs will have  cooled to become later types,  resulting in a more
``bottom-heavy'' luminosity function with respect to local estimates.

Additionally, this cooling should tend  to make early- to mid-L dwarfs
a    kinematically    younger    population    than    the    M-dwarfs
\citet[e.g.][]{seif10}.    However  the   high   velocity  dispersions
reported  by   many  kinematic   studies  suggest  ages   of  1--6~Gyr
\citep[e.g.][]{zapa07,fah09,seif10},  with a  well-established  age of
$\sim\!3$~Gyr  for  the  M-dwarfs \citep[e.g.][]{reid02}.   If  dwarfs
immediately below the  hydrogen-burning limit are indeed kinematically
younger (and  have a lower  velocity dispersion) than the  lowest mass
main sequence  dwarfs, then  we expect they  will be distributed  in a
thinner disk.  Yet cooler spectral types will contain a mixture of old
(high mass) dwarfs that have cooled and young (low mass) objects. This
population will  then be kinematically  older, have a  higher velocity
dispersion,  and reside  in  a  thicker disk  than  the warmest  brown
dwarfs.  Therefore we  expect to see a gradual  deviation in the trend
of scale  height with spectral type, since  the hydrogen-burning limit
does not  occur for a fixed  spectral type. With the  present data, we
find a scale height of $z_{scl}\!=\!\zscl\pm\dztot$~pc for a sample of
M8--T  dwarfs,  which is  comparable  to  estimates  for mid-M  dwarfs
\citep{juric08,boch10}  and  is   qualitatively  consistent  with  the
kinematic  results \citep[e.g.][]{fah09}.   However,  our estimate  is
somewhat lower  than the  extrapolation of the  trend of  scale height
with spectral type \citep[see Figure~10 of][]{juric08}. More data with
greatly improved  spectral typing is  needed to fully  constrain these
effects.

Our sample may contain dwarfs as  early as $\sim$M8, which is a direct
consequence of  the filter set.  These infrared  colors are determined
mostly by a  series of molecular absorption bands,  notably H$_2$O and
CH$_4$,  which  are  in  turn   used  to  define  the  spectral  types
\citep[e.g.][and  references therein]{kirk05}.   Therefore,  a cleaner
selection  and spectral  typing can  be  achieved by  using medium  or
narrow    bands     which    isolate    these     spectral    features
\citep[e.g.][]{jones94}.    For  example,   the   H$_2$O  feature   at
$\lambda\!=\!1.34~\mu$m  directly   maps  onto  effective  temperature
\citep{jones95} and  is relatively insensitive to  surface gravity and
metallicity  \citep{gor03,wilk04}.  Unfortunately  the  $J_{125}$- and
$H_{160}$-bands equally split the H$_2$O feature, diminishing the type
discrimination  of  the  $(J_{125}-H_{160})$  color.   Future  surveys
dedicated to  finding ultracool dwarfs  could take advantage  of these
molecular  features  for  robust  identification  and  classification.
Fortunately, WFC3 contains  a host of filters designed  to sample this
H$_2$O  absorption feature  \citep{lupie}, specifically  F127M, F139M,
and F153M.   Furthermore, the {\it James Webb  Space Telescope} (JWST)
and its  Near-Infrared Camera (NIRCam)  will be equipped  with similar
bandpasses,   but  with   a  significantly   larger   collecting  area
facilitating surveys at still larger heliocentric distances and search
for ultracool  dwarfs associated with other  Galactic components (such
as thick disk, halo, or bulge).

\acknowledgments We thank the anonymous Referee and Adam Burgasser for
many  insightful and  helpful comments  and suggestions.   Support for
this work was provided by  NASA through grant numbers 11772 (for RER),
11702 (for HY  and MRM), and 11359 (for SHC)  from the Space Telescope
Science  Institute,  which  is  operated  by AURA,  Inc.,  under  NASA
contract  NAS  5-26555.   RAW  acknowledges  support  from  NASA  JWST
Interdisciplinary Scientist grant NAG5-12469 from GSFC.

{\it Facilities:} \facility{HST (WFC3)}

\end{document}

%% file: tab1.tex
\begin{table*}
\caption{Survey Fields}
\label{fields}
\begin{tabular*}{0.98\textwidth}
  {@{\extracolsep{\fill}}lcccccccr}
\hline\hline

\multicolumn{1}{c}{Field} & \multicolumn{1}{c}{RA$^\dagger$} & \multicolumn{1}{c}{Dec$^\dagger$} & \multicolumn{1}{c}{$\ell^\dagger$} & \multicolumn{1}{c}{$b^\dagger$} & \multicolumn{1}{c}{$\Delta\Omega^\ddagger$} & \multicolumn{1}{c}{$J_{50}^*$} & \multicolumn{1}{c}{$A_J^{**}$} & \multicolumn{1}{c}{Optical}\\
\multicolumn{1}{c}{$ $} & \multicolumn{1}{c}{($^{\rm h}\;^{\rm m}\;^{\rm s}$)} & \multicolumn{1}{c}{($^\circ\;'\;"$)} & \multicolumn{1}{c}{(deg)} & \multicolumn{1}{c}{(deg)} & \multicolumn{1}{c}{($\square'$)} & \multicolumn{1}{c}{(mag)} & \multicolumn{1}{c}{(mag)} & \multicolumn{1}{c}{Band}\\
\hline
par0110$-$0222 & $01\;10\;09.45$ & $-02\;22\;23.0$ & 133.987232 & $-64.842182$ &  4.68 & 27.69 & 0.04 &   BOTH \\
par0213$+$1254 & $02\;13\;38.75$ & $+12\;54\;59.2$ & 152.018742 & $-45.261886$ &  4.69 & 27.02 & 0.11 & F600LP \\
cos0227$-$4101 & $02\;27\;56.91$ & $-41\;01\;34.4$ & 254.161369 & $-65.792729$ &  4.69 & 28.00 & 0.01 & F600LP \\
cos0240$-$1857 & $02\;40\;27.63$ & $-18\;57\;14.4$ & 200.649009 & $-63.686780$ &  4.69 & 27.95 & 0.03 & F600LP \\
           ERS & $03\;32\;23.03$ & $-27\;42\;50.2$ & 223.407959 & $-54.441403$ & 46.49 & 28.20 & 0.01 &  F606W \\
cos0439$-$5316 & $04\;39\;25.42$ & $-53\;16\;40.4$ & 261.334943 & $-40.946276$ &  4.69 & 28.21 & 0.00 & F600LP \\
par0539$-$6409 & $05\;39\;30.82$ & $-64\;09\;03.4$ & 273.650747 & $-32.015470$ &  4.72 & 26.45 & 0.05 &  F606W \\
par0553$-$6405 & $05\;53\;06.02$ & $-64\;05\;18.0$ & 273.525663 & $-30.535557$ &  4.76 & 27.01 & 0.04 &  F606W \\
par0623$-$6431 & $06\;23\;34.06$ & $-64\;31\;49.1$ & 274.232994 & $-27.264246$ &  4.68 & 26.32 & 0.05 &  F606W \\
par0623$-$6439 & $06\;23\;48.13$ & $-64\;39\;41.0$ & 274.382687 & $-27.253780$ &  4.71 & 26.94 & 0.05 &  F606W \\
par0637$-$7519 & $06\;37\;05.02$ & $-75\;18\;39.4$ & 286.419000 & $-27.078161$ &  6.92 & 26.98 & 0.09 &  F606W \\
par0750$+$2917 & $07\;50\;50.58$ & $+29\;16\;53.6$ & 191.358334 & $+24.960307$ &  4.81 & 27.43 & 0.04 &   BOTH \\
par0755$+$3043 & $07\;55\;57.08$ & $+30\;43\;10.9$ & 190.214896 & $+26.453597$ &  4.68 & 26.71 & 0.06 &  F606W \\
par0808$+$3945 & $08\;08\;21.38$ & $+39\;45\;25.3$ & 180.923544 & $+31.128501$ &  4.68 & 25.49 & 0.04 &  F606W \\
par0819$+$4911 & $08\;19\;19.04$ & $+49\;11\;05.4$ & 170.093673 & $+34.244836$ &  4.68 & 27.72 & 0.05 &  F606W \\
par0820$+$2332 & $08\;20\;03.41$ & $+23\;32\;05.0$ & 199.823952 & $+29.326292$ &  4.69 & 27.29 & 0.04 &  F606W \\
cos0846$+$7653 & $08\;46\;22.36$ & $+76\;53\;39.8$ & 136.607977 & $+32.760135$ &  4.68 & 28.35 & 0.02 & F600LP \\
par0905$+$0255 & $09\;05\;37.52$ & $+02\;55\;31.6$ & 226.848178 & $+30.961043$ &  4.68 & 27.03 & 0.03 &  F606W \\
par0909$-$0001 & $09\;09\;09.14$ & $-00\;01\;47.1$ & 230.318031 & $+30.194936$ &  4.68 & 27.36 & 0.03 &  F606W \\
par0914$+$2821 & $09\;14\;16.82$ & $+28\;21\;44.6$ & 198.147026 & $+42.355984$ &  4.68 & 27.50 & 0.02 &  F606W \\
par0921$+$4505 & $09\;21\;38.15$ & $+45\;05\;08.0$ & 175.142310 & $+44.900120$ &  4.68 & 27.16 & 0.02 &  F606W \\
par0925$+$4425 & $09\;25\;32.15$ & $+44\;25\;31.8$ & 175.989494 & $+45.648179$ &  4.69 & 27.85 & 0.01 & F600LP \\
par0925$+$4000 & $09\;25\;35.45$ & $+40\;00\;13.0$ & 182.321062 & $+45.878836$ &  4.68 & 27.49 & 0.01 &  F606W \\
par1030$+$3803 & $10\;30\;52.52$ & $+38\;03\;24.5$ & 183.565827 & $+58.665121$ &  4.68 & 27.62 & 0.01 &  F606W \\
cos1131$+$3117 & $11\;31\;29.93$ & $+31\;17\;21.8$ & 194.732083 & $+72.094827$ &  4.69 & 27.98 & 0.02 & F600LP \\
par1151$+$5441 & $11\;51\;49.26$ & $+54\;40\;59.8$ & 140.435934 & $+60.372625$ &  4.71 & 27.70 & 0.01 &  F606W \\
par1152$+$0056 & $11\;52\;43.92$ & $+00\;55\;51.2$ & 272.228218 & $+60.255125$ &  4.68 & 27.78 & 0.02 &  F606W \\
par1209$+$4543 & $12\;09\;24.82$ & $+45\;43\;26.1$ & 144.367666 & $+69.615667$ &  4.70 & 28.09 & 0.01 &   BOTH \\
par1244$+$3356 & $12\;44\;45.21$ & $+33\;56\;05.1$ & 134.455667 & $+83.043441$ &  4.68 & 28.13 & 0.01 &  F606W \\
par1301$-$0000 & $13\;01\;16.61$ & $-00\;00\;27.0$ & 308.312235 & $+62.761347$ &  4.68 & 27.38 & 0.02 & F600LP \\
par1336$-$0027 & $13\;36\;48.75$ & $-00\;27\;57.9$ & 326.341678 & $+60.326874$ &  4.68 & 27.86 & 0.03 & F600LP \\
par1340$+$4123 & $13\;40\;31.87$ & $+41\;23\;03.3$ &  90.813506 & $+72.543346$ &  4.68 & 28.17 & 0.01 & F600LP \\
par1436$+$5043 & $14\;36\;56.72$ & $+50\;42\;58.6$ &  89.753150 & $+59.068438$ &  4.69 & 28.19 & 0.01 &  F606W \\
par1524$+$0954 & $15\;24\;10.17$ & $+09\;54\;19.8$ &  14.751734 & $+50.137021$ &  4.68 & 27.64 & 0.04 & F600LP \\
par1631$+$3736 & $16\;31\;34.28$ & $+37\;36\;21.4$ &  60.300146 & $+43.026120$ &  4.68 & 27.83 & 0.01 &  F606W \\
par1632$+$3733 & $16\;32\;18.38$ & $+37\;33\;24.3$ &  60.246932 & $+42.877428$ &  4.68 & 27.64 & 0.01 &  F606W \\
cos2057$-$4412 & $20\;57\;22.01$ & $-44\;12\;26.9$ & 356.582832 & $-40.624781$ &  4.68 & 27.45 & 0.03 & F600LP \\
cos2202$+$1851 & $22\;02\;48.68$ & $+18\;50\;58.5$ &  76.653695 & $-28.493144$ &  4.92 & 28.10 & 0.06 & F600LP \\
par2345$-$0054 & $23\;45\;02.34$ & $-00\;54\;11.0$ &  88.894697 & $-59.313985$ &  4.68 & 27.83 & 0.03 & F600LP \\
cos2350$-$4331 & $23\;50\;36.39$ & $-43\;31\;30.3$ & 335.844576 & $-69.509649$ &  4.68 & 28.18 & 0.01 & F600LP \\
\hline
\multicolumn{9}{l}{$^\dagger$Coordinates refer to the field center in the J2000 epoch.}\\
\multicolumn{9}{l}{$^{\ddagger}$Solid angle in arcmin$^2$.}\\
\multicolumn{9}{l}{$^*$The approximate 50\% completeness limit.}\\
\multicolumn{9}{l}{$^{**}$The $J$-band extinction from \citet{sfd}.}
\end{tabular*}
\end{table*}

%% file: tab2.tex
\begin{table}
\caption{SpeX Catalog$^\dagger$}
\label{spexcat}
\begin{tabular*}{0.48\textwidth}
  {@{\extracolsep{\fill}}lr}
\hline\hline
\multicolumn{1}{c}{Reference} & \multicolumn{1}{c}{Number of}\\
           \multicolumn{1}{c}{$ $}       & \multicolumn{1}{c}{Citations}\\
\hline
A. Burgasser (unpublished) & 389 \\
          \citet{burg2010} & 116 \\
         \citet{burg2004a} &  87 \\
          \citet{chiu2006} &  51 \\
         \citet{burg2008a} &  45 \\
         \citet{burg2006a} &  19 \\
        \citet{muench2007} &  17 \\
           \citet{loop07a} &  14 \\
         \citet{burg2006b} &  12 \\
           \citet{burg07a} &   8 \\
            \citet{sieg07} &   7 \\
           \citet{burg07b} &   6 \\
        \citet{burgmcel06} &   6 \\
          \citet{shepcush} &   6 \\
            \citet{cruz04} &   5 \\
        \citet{burgkirk06} &   4 \\
            \citet{loop08} &   3 \\
         \citet{burg2004b} &   3 \\
           \citet{loop07b} &   3 \\
              \citet{mb06} &   3 \\
            \citet{reid06} &   3 \\
            \citet{muno06} &   2 \\
         \citet{burg2008b} &   2 \\
          \citet{burg2009} &   1 \\
        \citet{luhman2007} &   1 \\
          \citet{kirk2006} &   1 \\
           \citet{burg07c} &   1 \\
           \citet{burg07d} &   1 \\
          \citet{liebburg} &   1 \\
\hline
\multicolumn{2}{l}{$^\dagger$Compiled A.~Burgasser and distributed at}\\
\multicolumn{2}{l}{http://web.mit.edu/ajb/www/browndwarfs/spexprism/}
\end{tabular*}
\end{table}

%% file: tab3.tex
\begin{table*}
\caption{Ultracool Dwarf Candidates}
\label{cands}
\begin{tabular*}{0.98\textwidth}
  {@{\extracolsep{\fill}}lcccccccc}
\hline\hline
\multicolumn{1}{c}{ID} & \multicolumn{1}{c}{RA$^\dagger$} & \multicolumn{1}{c}{Dec$^\dagger$} & \multicolumn{1}{c}{($V_{606}-Y_{098}$)} & \multicolumn{1}{c}{($I_{600}-Y_{098}$)} & \multicolumn{1}{c}{($Y_{098}-J_{125}$)} & \multicolumn{1}{c}{($J_{125}-H_{160}$)} & \multicolumn{1}{c}{$J_{125}$} & \multicolumn{1}{c}{SpT$^{ddagger}$} \\
\multicolumn{1}{c}{$ $} & \multicolumn{1}{c}{($^{\rm h}\;^{\rm m}\;^{\rm s}$)} & \multicolumn{1}{c}{($^\circ\;'\;"$)} & \multicolumn{1}{c}{(mag)} & \multicolumn{1}{c}{(mag)} & \multicolumn{1}{c}{(mag)} & \multicolumn{1}{c}{(mag)} & \multicolumn{1}{c}{(mag)} & \multicolumn{1}{c}{$ $}\\
\hline
 1 & $06\;23\;27.31$ & $-64\;31\;22.0$ &  $4.73\pm0.34$ &       \nodata & $1.06\pm0.01$ & $ 0.35\pm0.01$ & $20.88\pm0.00$ &   L \\
 2 & $09\;25\;32.75$ & $+44\;24\;44.5$ &        \nodata & $0.59\pm0.05$ & $1.16\pm0.02$ & $ 0.31\pm0.01$ & $22.53\pm0.01$ &   L \\
 3 & $22\;02\;45.31$ & $+18\;50\;53.2$ &        \nodata & $0.62\pm0.08$ & $1.18\pm0.03$ & $ 0.48\pm0.02$ & $23.19\pm0.01$ &   L \\
 4 & $02\;13\;33.79$ & $+12\;54\;11.4$ &        \nodata & $1.41\pm0.37$ & $1.13\pm0.11$ & $ 0.35\pm0.05$ & $23.36\pm0.04$ &   L \\
 5 & $13\;36\;46.69$ & $-00\;28\;35.3$ &        \nodata & $1.16\pm0.14$ & $0.84\pm0.03$ & $ 0.50\pm0.03$ & $23.41\pm0.02$ & MLT \\
 6 & $13\;01\;13.05$ & $+00\;00\;09.0$ &        \nodata & $1.45\pm0.24$ & $0.83\pm0.06$ & $ 0.43\pm0.04$ & $23.43\pm0.03$ & MLT \\
 7 & $16\;32\;17.16$ & $+37\;33\;32.7$ &  $2.36\pm0.33$ &       \nodata & $0.89\pm0.04$ & $ 0.43\pm0.03$ & $23.57\pm0.02$ & MLT \\
 8 & $09\;25\;37.82$ & $+40\;01\;03.9$ &  $2.77\pm0.90$ &       \nodata & $0.86\pm0.07$ & $ 0.19\pm0.05$ & $23.76\pm0.03$ & MLT \\
 9 & $08\;46\;16.63$ & $+76\;53\;12.5$ &        \nodata & $0.72\pm0.34$ & $1.34\pm0.10$ & $ 0.64\pm0.04$ & $24.21\pm0.03$ &   L \\
10 & $16\;32\;21.30$ & $+37\;32\;52.1$ &  $2.53\pm0.77$ &       \nodata & $0.93\pm0.08$ & $ 0.39\pm0.04$ & $24.21\pm0.03$ & MLT \\
11 & $15\;24\;08.81$ & $+09\;55\;06.2$ &        \nodata & $1.84\pm0.70$ & $0.87\pm0.08$ & $ 0.19\pm0.07$ & $24.65\pm0.05$ & MLT \\
12 & $16\;31\;32.80$ & $+37\;35\;53.6$ &      $>\!3.12$ &       \nodata & $0.86\pm0.13$ & $ 0.19\pm0.10$ & $24.72\pm0.06$ & MLT \\
13 & $04\;39\;21.53$ & $-53\;16\;52.0$ &        \nodata & $0.65\pm0.30$ & $1.16\pm0.11$ & $ 0.49\pm0.04$ & $24.73\pm0.03$ &   L \\
14 & $06\;23\;39.94$ & $-64\;30\;58.3$ &        \nodata &       \nodata & $0.85\pm0.29$ & $ 0.35\pm0.19$ & $24.86\pm0.12$ & MLT \\
15 & $09\;14\;22.12$ & $+28\;21\;34.6$ &      $>\!3.01$ &       \nodata & $0.76\pm0.10$ & $ 0.18\pm0.07$ & $24.93\pm0.05$ & MLT \\
16 & $04\;39\;26.10$ & $-53\;16\;01.1$ &        \nodata &     $>\!2.92$ & $0.93\pm0.11$ & $-0.28\pm0.07$ & $25.08\pm0.03$ &   T \\
17 & $08\;19\;19.00$ & $+49\;11\;02.3$ &      $>\!2.61$ &       \nodata & $0.81\pm0.21$ & $ 0.25\pm0.18$ & $25.28\pm0.11$ & MLT \\
\hline
\multicolumn{9}{l}{$^\dagger$Coordinates refer to the J2000 equinox.}\\
\multicolumn{9}{l}{$^\ddagger$Spectral types based on equations~(\ref{mltdwarf}), (\ref{ldwarf}), and (\ref{tdwarf}).}
\end{tabular*}
\end{table*}